\documentclass[sort&compress,12pt,draftcls,onecolumn]{IEEEtran}

\ifCLASSINFOpdf
\usepackage[pdftex]{graphicx}

\else

\fi

\usepackage[cmex10]{amsmath}
\usepackage{amssymb}   
\usepackage{amsxtra}
\usepackage{amscd}
\usepackage{booktabs}
\usepackage{amsthm}
\usepackage{textcomp}
\usepackage{graphicx}  
\usepackage{balance}
\usepackage{array}  
\usepackage{multirow}  
\usepackage{amsmath}
\usepackage{cite}
\usepackage{times}
\usepackage{blindtext}
\usepackage{array}
\usepackage{tabularx}
\usepackage[utf8]{inputenc}
\usepackage{makecell}

\usepackage{amsmath}

\usepackage{enumitem}
\usepackage{wasysym}
\usepackage{framed}
\usepackage{pgfgantt,rotating}
\usepackage{subfloat}
   \usepackage[flushleft]{threeparttable}

\begin{document}
	
	\title{{\huge Channel Modeling in RIS-Empowered Wireless Communications}}

	\author{ Ibrahim~Yildirim,~\IEEEmembership{Student Member,~IEEE} \\ and Ertugrul~Basar,~\IEEEmembership{Senior Member,~IEEE}
		\vspace*{-0.25cm}
		\thanks{I. Yildirim and E. Basar are with the Communications Research and Innovation Laboratory (CoreLab), Department of Electrical and Electronics Engineering, Ko\c{c} University, Sariyer 34450, Istanbul, Turkey. e-mail: ebasar@ku.edu.tr}
		\thanks{I. Yildirim is also with the Faculty of Electrical and Electronics Engineering, Istanbul Technical University, Istanbul 34469, Turkey. e-mail: yildirimib@itu.edu.tr}}

	\maketitle

	\begin{abstract}
		One of the most critical aspects of enabling next-generation wireless technologies is developing an accurate and consistent channel model to be validated effectively with the help of real-world measurements. From this point of view, remarkable research has recently been conducted to model propagation channels involving the modification of the wireless propagation environment through the inclusion of reconfigurable intelligent surfaces (RISs). This study mainly aims to present a vision on channel modeling strategies for the RIS-empowered communications systems considering the state-of-the-art channel and propagation modeling efforts in the literature. Moreover, it is also desired to draw attention to open-source and standard-compliant physical channel modeling efforts to provide comprehensive insights regarding the practical use-cases of RISs in future wireless networks.
	\end{abstract}
	


	\IEEEpeerreviewmaketitle
	
\section{Introduction}

Despite the vast drastic expectations, fifth-generation (5G) wireless communication networks can be considered as an evolution of fourth-generation (4G) wireless networks by providing a more adaptive and flexible structure in terms of adapted physical layer technologies. In sixth-generation (6G) wireless networks, which are expected to be rolled out in 2030s, it is an undeniable fact that there should be paradigm-shifting developments, especially in the physical layer, in order to meet the increasing massive demand in data consumption \cite{Samsung_2020}. Within this context, millimeter wave (mmWave) and Terahertz (THz) communications, massive multiple-input multiple-output systems (MIMO), cell-free networks, high-altitude platform stations, integrated space and terrestrial networks, and reconfigurable intelligent surfaces (RISs) can be considered as promising candidate technologies for 6G systems in order to satisfy broadband connectivity by keeping new use-cases with very high mobility and extreme capacity \cite{Rajatheva_6G}.

\begin{figure}
\centering
\includegraphics[width=13cm]{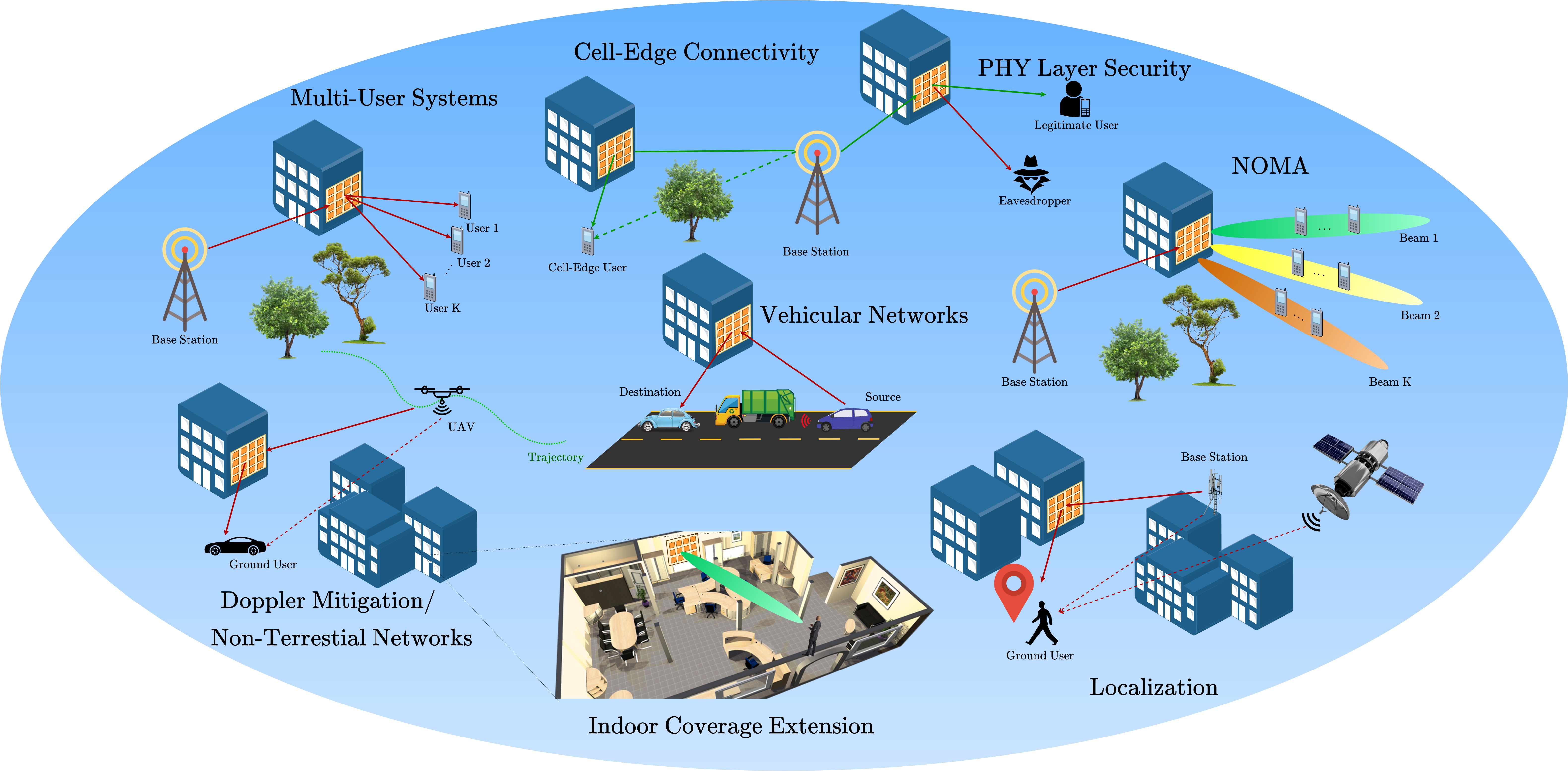}
\caption{Miscellaneous RIS-assisted transmission scenarios and applications that  are likely to emerge in future wireless communication networks. Source: Reprinted and adapted with permission from Basar and Yildirim \cite{SimRIS_Mag}. Copyright 2021, IEEE.
\label{fig1-1}}
\end{figure}

RIS-empowered communication  can be regarded as a paradigm-shifting revolution that provides operators software-based and dynamic control capability over the wireless propagation channel \cite{Basar_Access_2019}. An RIS, which consists of large number of nearly passive reflecting elements, can intelligently direct the beams to the desired users and ensure reliable communication by creating virtual line-of-sight (LOS) links even when the direct link between terminals is blocked. Due to these unprecedented contributions of RISs, researchers have shown remarkable interest in RIS-empowered communication systems, and plentiful applications that are candidates for use in next-generation communication systems have been explored \cite{Wu_Tutorial,Basar_2019_LIS_2,Yildirim_hybrid,Yildirim_multiRIS,Arslan_2021}. 
The various use-cases of RISs, including multi-user systems, physical (PHY) layer security, indoor-coverage extension, Doppler mitigation, vehicular and non-terrestrial networks, localization and sensing, and non-orthogonal multiple access (NOMA), are illustrated in Fig. \ref{fig1-1}. 
Despite these rich RIS use-cases, there is no strong consensus to identify the killer applications that effectively exploit the potential of RISs to control the transmission medium with intelligent reflections for enhanced end-to-end system performance \cite{SimRIS_Mag}. The first step in clearing this ambiguity is to form a unified and physical RIS-assisted channel model that can be adapted to different use-cases and operating frequencies, taking into account the physical characteristics of RISs. Since intelligent reflection is the art of manipulating the channel characteristics in a brilliant and dynamic way, modeling the RIS-assisted transmission link is essential to provide detailed insights into the practical use-cases. This work mainly aims to present a vision on channel modeling strategies for the RIS-empowered communications systems considering the state-of-the-art channel and propagation modeling efforts in the literature. Another objective of the work is to draw attention to open-source and standard-compliant physical channel modeling efforts to provide comprehensive insights regarding the practical use-cases of RISs in future wireless networks.

The rest of the work is organized as follows: Section 1.2 presents a  general framework on channel modeling strategies for the RIS-empowered communications systems considering the state-of-the-art channel modeling efforts in the literature. Section 1.3 and 1.4 provide a framework on the cluster-based statistical channel model is summarized for sub-6 GHz and mmWave bands, respectively. Section 1.5 introduces the open-source \textit{SimRIS Channel Simulator}  package that considers a proposed narrowband channel model for RIS-empowered communication systems. Section 1.6 presents numerical results obtained by using \textit{SimRIS Channel Simulator}  and the work is concluded in Section 1.7.


\section{A General Perspective on RIS Channel Modeling}
\begin{figure}
\centering
\includegraphics[width=1\columnwidth]{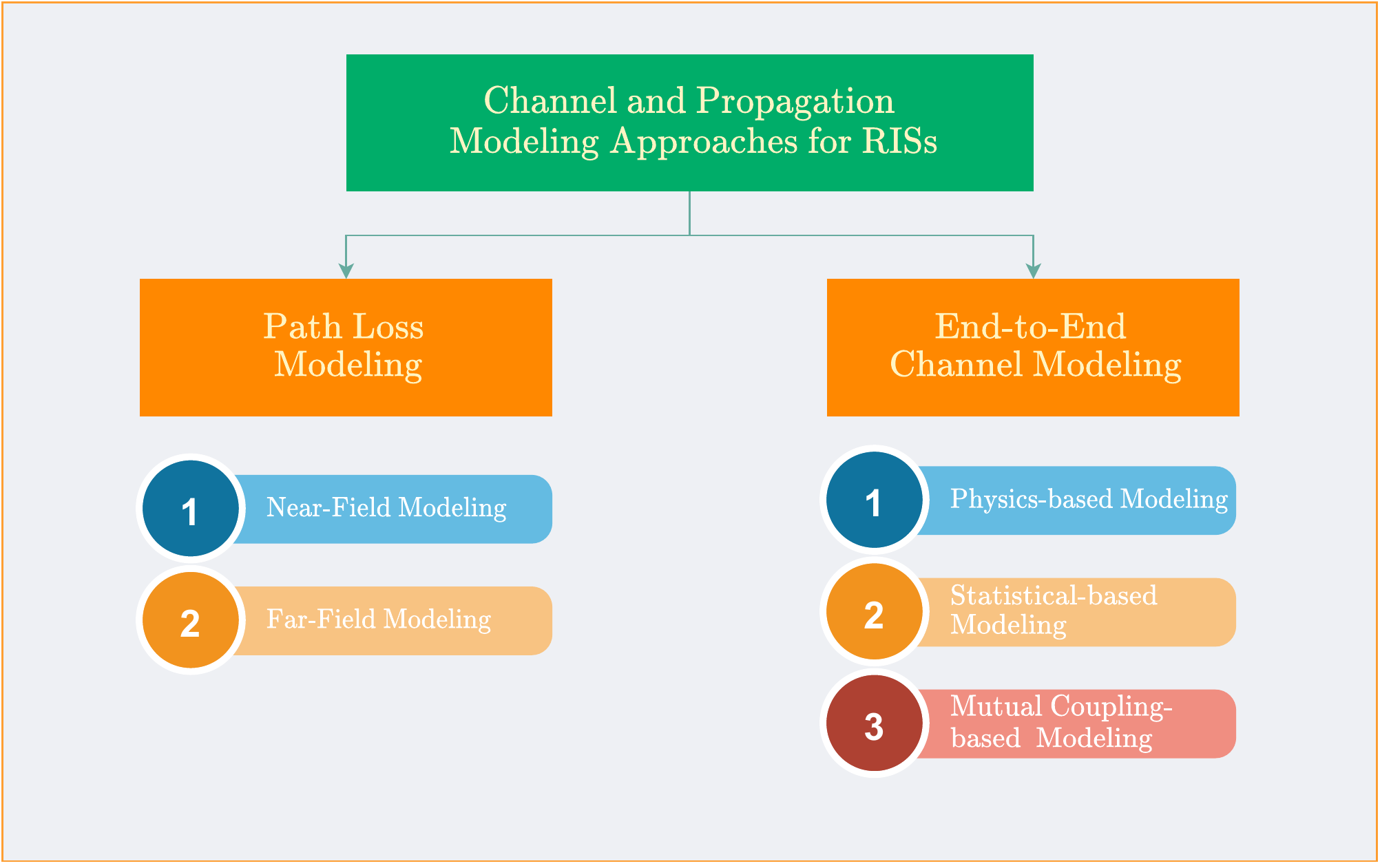}
\caption{Classification of the channel and propagation modeling approaches based on the existing RIS literature. \label{fig2-1}}
\end{figure}

One of the most important aspects of enabling next-generation wireless technologies is the development of an accurate and consistent channel model to be validated effectively with the help of real-world measurements. From this point of view, remarkable research has recently been conducted to model propagation channels involving the modification of the wireless propagation environment through the inclusion of RISs. While many studies in the literature have modeled the path loss for RIS-assisted communication systems, modeling the end-to-end channel has also recently attracted significant attention. The near-field and far-field effects of RISs should also be considered when modeling the total path loss of RIS-empowered systems. Another critical point is the methodology followed when modeling the end-to-end channel. From this perspective, while many researchers follow physics-based channel modeling strategies by examining the physical properties of reflection and scattering effects, statistical channel modeling strategies have also been followed by considering the statistical properties of scatterers and clusters in the environment. Further, electromagnetic (EM) compatible channel models have also been proposed that take into account EM properties, such as mutual coupling among the sub-wavelength unit cells of the RIS. By considering the existing RIS-assisted channel modeling studies, the channel and propagation modeling approaches can be mainly classified as in Fig. \ref{fig2-1}.  
Since there is no obvious distinction between certain approaches in this classification, channel modeling studies in which more than one approach is used can provide a broader perspective.

Initial studies on RIS propagation modeling have predominantly focused on the path-loss and scattered power characterization by an RIS in RIS-assisted communication systems \cite{Renzo_Analy2020,Garcia_2019,Ellingson,Ozdogan_2020,Tang_2020,Danufane_2021,Emilpower_2020}. In \cite{Renzo_Analy2020}, the authors characterize the path-loss in near-field and far-field of RISs by using the general scalar theory of diffraction and the Huygens-Fresnel principle. Moreover, the power reflected from an RIS is obtained as a function of the distance between the transmitter (Tx)/receiver (Rx) and the RIS, the size of the RIS, and the phase shifts induced by the RIS. By modeling RISs as a sheet of EM material of negligible thickness, the conditions under which an RIS acts as an anomalous mirror are identified. In \cite{Garcia_2019}, the authors characterize the path-loss using physics-based approaches and exploited the antenna theory on the calculation of the electric field both in the far-field and near-field of a finite-size RIS. While it is shown that an RIS can act as an anomalous mirror in the near field of the array, the received power expression is not formulated analytically for varying distances. The author in \cite{Ellingson} calculates the path-loss as a function of RIS size, link geometry and practical gain of RIS elements by considering a passive reflectarray-type RIS design under the far-field and near-field cases. The practical power scaling laws are introduced in \cite{Ozdogan_2020} for the far-field region of array using the physical optic techniques. The authors  also clarify why the surface comprises a large number of reflecting elements that individually act as diffuse scatterers but can jointly beamform the signal in the desired direction with a specific beamwidth.  In \cite{Tang_2020}, the free-space path-loss models for RIS-enabled communication systems are characterized with physics and EM-based approaches by considering the distances from the Tx/Rx to the RIS, the size of the RIS, the near-field/far-field effects of the RIS, and the radiation patterns of antennas and unit cells. The proposed models are validated through extensive computer simulation results and experimental measurements using a specifically manufactured RIS. In \cite{Danufane_2021}, the authors leverage the vector generalization of Green’s theorem and characterize the free space path-loss of the RIS-assisted communications by using physical optic methods in order to overcome the limitation of geometric optics. The analyses are conducted for two-dimensional (2D) homogenized metasurfaces that can operate in reflection or refraction mode under both far-field and near-field cases. Furthermore, the power scaling laws for asymptotically large RISs are studied in \cite{Emilpower_2020} using a deterministic propagation model and observed that the asymptotic limit could only be achieved in the near-field case.

While aforementioned studies characterize the path loss by using physics and EM-based approaches,  in \cite{Najafi_2021}, a physics-based end-to-end channel model for RIS-empowered communication systems is introduced and a scalable optimization framework for large RISs is presented. This model includes 
the impact of all RIS tiles, the transmission modes of all tiles, and the incident, reflection, and polarization angles. Furthermore, each tile is modeled as an anomalous reflector, and physical optics-based analyses are conducted by adopting the concepts from the radar literature under the far-field conditions.

In \cite{Gradoni_2021}, the authors propose a physics and EM-compliant end-to-end channel model for the RIS-assisted communication systems by considering the mutual coupling among the sub-wavelength unit cells of the RIS.  This mutual coupling and unit cell aware model can be applied to an RIS consisting of closely-spaced scattering elements controlled by tunable impedances. Inspiring from the impedance-based channel model in \cite{Gradoni_2021}, an analytical framework and a iterative algorithm are presented to obtain the end-to-end received power in \cite{Qian_2021}. In \cite{Abrardo_2021}, a multi-user multiple-input multiple-output (MIMO) interference network is investigated in the presence of multiple RISs by using a circuit-based model for the transmitters, receivers, and RISs. The mutual coupling between impedance-controlled thin dipoles and the impact of the tuning elements are also investigated. Additionally, a provably convergent optimization algorithm is proposed to maximize the sum-rate of RIS-assisted MIMO interference channels by assessing the mutual coupling among closely-spaced scattering elements.

Recently, there have been a rapid and growing interest on stochastic non-stationary channel modeling activities for RIS-empowered wireless systems \cite{Sun1_2021,Sun2_2021,GSun_2021,Jiang_2021, Xiong_2021}. Within this context, in \cite{Sun1_2021}, a three-dimensional (3D) geometry-based stochastic channel model (GBSM) is introduced for a massive MIMO communication system in the presence of an RIS. The authors split the end-to-end channel into the sub-channels and characterize the large and small scale fading, respectively. In \cite{Sun2_2021}, practical phase shifts are considered and the reflection phases of the RIS are represented using two-bit quantization set in addition to the work in \cite{Sun1_2021}. However, these two studies do not consider the practical deployment and physical properties of RISs. The study in \cite{GSun_2021} proposes a geometric RIS-assisted MIMO channel model for fixed-to-mobile communications based on a 3D cylinder model. While this study examines the non-stationary channels in which the receiver moves, the characteristic features of the scatterers in the environment are ignored, and analyses are merely conducted under sub-6 GHz frequency bands. Recently, the authors in \cite{Jiang_2021} propose a general wideband non-stationary channel model for RIS-empowered communication systems operating at sub-6 GHz bands. By splitting the RIS-assisted MIMO channel model into subchannels, equivalent end-to-end channel models based on these subchannels are characterized under time-varying characteristics of MIMO systems in the presence of an RIS. Although the authors consider the physical characteristics of the wideband subchannel model between the mobile Tx and RIS, and RIS and mobile Rx for different propagation delays, LOS links between the Tx and Rx are ignored, and frequency bands above 6 GHz are not considered. More recently, a non-stationary 3D wideband GBSM for RIS-empowered MIMO communication systems are introduced in \cite{Xiong_2021} by including propagation characteristics of RISs, such as unit number and size, relative locations among Tx, RIS, and Rx and RIS configuration. Nevertheless, this study does not consider a direct link between the Tx and the mobile Rx due to the blockage, and the applicability of this model is limited with sub-6 GHz frequency bands in an outdoor environment. 

Against this background, due to the spectrum shortage of sub-6 GHz frequency bands, it is unavoidable to migrate to the mmWave and THz bands in future wireless networks. Since transmission at higher frequencies makes the signals more vulnerable to blockage and interference, the signal attenuation and blockage prevent mmWaves from reaching long distances. Especially in mmWave frequencies, using an RIS in the transmission can enable additional transmission paths when the direct link between the Tx and Rx is blocked or not sufficiently robust. Therefore, modeling a physical open-source, and widely applicable mmWave channel for the RIS-assisted systems in indoor and outdoor environments is of great importance to shed light on realistic use-cases of RISs in future wireless networks. From this point of view, a unified narrowband channel model for RIS-assisted systems both in indoor and outdoor environments is introduced in \cite{SimRIS_Latincom,SimRIS_TCOM,SimRIS_Mag} by including the 5G mmWave channel model with a random number of clusters/scatterers and the characteristics of the RIS. Moreover, a physical channel model for RIS-assisted systems is proposed in \cite{Kilinc_2021} by considering the currently used technical specifications on sub-6 GHz bands.

\begin{table}
\centering \scriptsize 
\caption{An overview of RIS channel modeling studies and their main contributions.\label{tab1-1}}{%
\begin{tabular}{||c|p{0.8\textwidth}||}
\hline
\textbf{Work}     &\hspace{1.65in} {\textbf{Contributions}}   \\
\hline
\hline
\cite{Renzo_Analy2020} &\makecell{$\bullet$ Path-loss in near-field and far-field of RISs is characterized by using Huygens-Fresnel principle \\  and physical optics-based methods\\ $\bullet$ RISs are modeled as a sheet of EM material of negligible thickness } \\
\hline
\cite{Garcia_2019} &\makecell{$\bullet$ Path-loss is characterized with physics-based approach \\ $\bullet$ Electric field of a finite-size RIS is computed  both in its far-field and near-field}    \\
\hline
\cite{Ellingson} &\makecell{$\bullet$ Path-loss is calculated for a passive reflectarray-type RIS \\ $\bullet$
The path-loss is characterized as a function of RIS size, link geometry, and practical gain of RIS elements} \\
\hline
\cite{Ozdogan_2020} &\makecell{$\bullet$ Practical power scaling laws are introduced for the far-field region of \\array using the physical optic principles} \\
\hline
\cite{Tang_2020} &\makecell{$\bullet$ Free-space path-loss with physic and EM-based approaches is characterized by considering \\  the near-field/far-field effects of the RIS\\$\bullet$ It is validated through experimental measurements using manufactured RIS} \\
\hline
\cite{Danufane_2021} &\makecell{$\bullet$ Path-loss is characterized as a function of the transmission distance and the RIS size by using \\  physical optic methods and Huygens-Fresnel principle\\ 
$\bullet$ 2D homogenized metasurfaces that can operate in reflection or refraction mode are considered \\  under both far-field and near-field case} \\
\hline
\cite{Emilpower_2020} &\makecell{$\bullet$ Power scaling laws for asymptotically large RISs are analysed using a deterministic propagation model \\ $\bullet$ It is shown that the asymptotic limit can only be achieved in the near-field } \\
\hline
\cite{Najafi_2021} &\makecell{$\bullet$ End-to-end channel model is characterized including the impact of the physical parameters of an RIS \\ $\bullet$ Physical optics-based analysis are conducted by adopting the concepts from the radar literature \\ $\bullet$ Each tile is modeled as an anomalous reflector } \\
\hline
\cite{Gradoni_2021,Qian_2021,Abrardo_2021} &\makecell{$\bullet$ EM-compliant communication model is characterized by considering the mutual coupling \\ among the RIS elements \\
$\bullet$ The entire system is represented with a circuit-based model for all terminals \\
$\bullet$ The mutual coupling between impedance-controlled thin dipoles and \\ the impact of the tuning elements are investigated} \\
\hline
\cite{Sun1_2021,Sun2_2021,GSun_2021,Jiang_2021, Xiong_2021} &\makecell{$\bullet$ A statistical geometry-based non-stationary channel models are derived for the RIS-empowered MIMO systems\\
 $\bullet$ The various physical properties of RIS are considered, such as unit number and size, \\  relative locations among Tx, RIS, and Rx} \\
\hline
\cite{SimRIS_Latincom,SimRIS_TCOM,SimRIS_Mag,Kilinc_2021} &\makecell{$\bullet$ Comprehensive channel modeling efforts are introduced by considering 3D channel models for \\ practical deployments and physical RIS characteristics\\
$\bullet$ Open-source and widely applicable \textit{SimRIS Channel Simulator}  is introduced\\
$\bullet$ Technical specifications on sub-6 GHz and mmWave bands are considered} \\
\hline
\hline
\end{tabular}}{}
\end{table}

An overview of the aforementioned channel modeling studies and brief explanations of their main contributions and characteristics can be seen in Table~\ref{tab1-1} \cite{Renzo_Analy2020,Garcia_2019,Ellingson,Ozdogan_2020,Tang_2020,Danufane_2021,Emilpower_2020,Najafi_2021,Gradoni_2021,Qian_2021,Abrardo_2021,Sun1_2021,Sun2_2021,GSun_2021,Jiang_2021,Xiong_2021,SimRIS_Latincom,SimRIS_TCOM,SimRIS_Mag}.

\section{Physical Channel Modeling for RIS-Empowered Systems at mmWave Bands}

   This section introduces a unified signal/channel model for RIS-assisted 6G communication systems operating under mmWave frequencies. This model is generic and can be applied to indoor and outdoor environments and operating frequencies. Considering the promising potential of RIS-assisted systems in future wireless networks at mmWave frequencies, a framework on the clustered statistical MIMO model is proposed by considering 3GPP standardization \cite{3GPP_5G}, while a generalization is possible. Because of their unique functionality, channel modeling methodology for RISs is different from other counterpart technologies, such as relaying. Although the major steps for generation RIS-assisted channel modeling are included in this work, interested readers are referred to \cite{SimRIS_Latincom,SimRIS_TCOM} for the comprehensive view of technical details and statistical characteristics of the channel parameters.

\begin{figure*}[!t]
	\begin{center}
		\includegraphics[width=1\columnwidth]{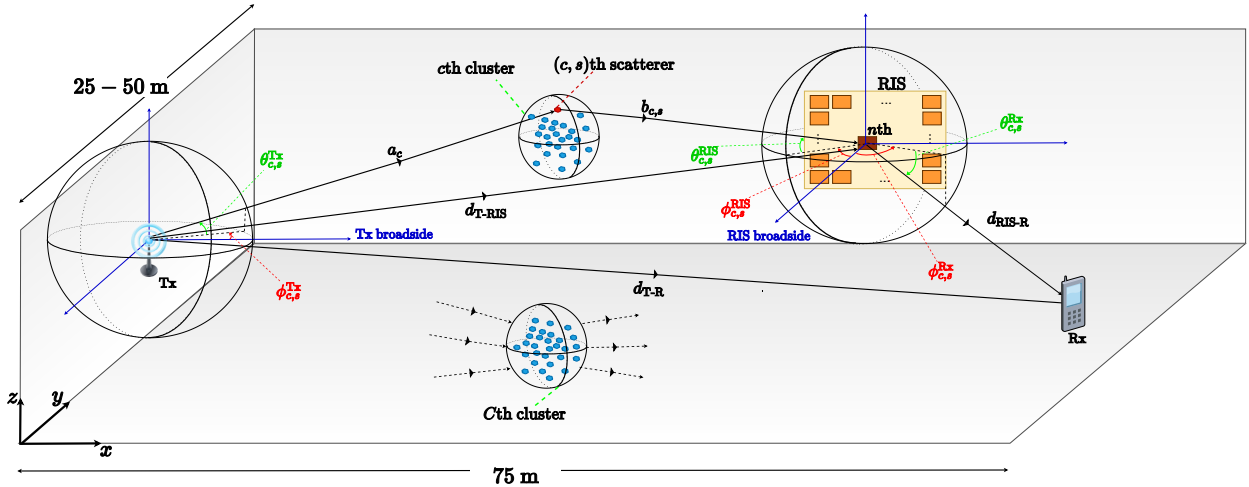}
		\vspace*{-0.3cm}\caption{Generic InH Indoor Office environment with $C$ clusters between Tx-RIS and an RIS mounted in the $xz$ plane (side wall). Source: Reprintedand adapted with permission from Basar, Yildirim and Kilinc \cite{SimRIS_TCOM}. Copyright 2021, IEEE.}\vspace*{-0.8cm}
		\label{fig3-2}
	\end{center}
\end{figure*}

       In the following, our channel modeling methodology is explained to generate Tx-RIS, RIS-Rx, and Tx-Rx subchannels for indoors and outdoors. It is assumed that the Tx lies on the $ yz $ plane, while the RIS lies either on the $ xz $ plane (Scenario 1 - side wall) or $ yz $ plane (Scenario 2 - opposite wall) for indoor and outdoor environments. In Fig. \ref{fig3-2}, the considered 3D geometry for a typical large indoor office is given as a reference for Scenario 1 (side-wall). For outdoor environments, this 3D geometry can be easily extended by modifying certain system parameters.
    
The existing interacting objects (scatterers) are assumed to be grouped under $C$ clusters, each having $S_c$ sub-rays for $c=1,\ldots,C$, that is $M=\sum_{c=1}^{C}S_c$. Therefore, the vector of Tx-RIS channel coefficients $\mathbf{h} \in \mathbb{C}^{N\times 1}$ can be obtained for a clustered model by considering array responses and path attenuations:
\begin{equation}\label{eq:13}
\mathbf{h}=
\gamma \sum\limits_{c=1}^{C} \sum\limits_{s=1}^{S_c} \beta_{c,s} \sqrt{G_e(\theta_{c,s}^{\text{RIS}})L_{c,s}^{\text{RIS}}} \,\, \mathbf{a} ( \phi_{c,s}^{\text{RIS}}, \theta_{c,s}^{\text{RIS}} ) + \mathbf{h}_{\text{LOS}}
\end{equation}
where $ \gamma= \sqrt{\frac{1}{\sum\nolimits_{c=1}^{C} S_c}}$ is a normalization factor, $\mathbf{h}_{\text{LOS}}$ is the LOS component, $\beta_{c,s} \sim \mathcal{CN} (0,1)$ and $L_{c,s}^{\text{RIS}}$ respectively stand for the complex path gain and attenuation associated with the $(c,s)$th propagation path, and $G_e(\theta_{c,s}^{\text{RIS}})$ is the RIS element pattern \cite{Nayeri} in the direction of the $(c,s)$th scatterer. Here, $\mathcal{CN} (0,\sigma^2)$ denotes complex Gaussian distribution with zero mean and $\sigma^2$ variance, and $\mathbf{a} ( \phi_{c,s}^{\text{RIS}}, \theta_{c,s}^{\text{RIS}} ) \in \mathbb{C}^{N\times 1}$ is the array response vector of the RIS for the considered azimuth ($ \phi_{c,s}^{\text{RIS}}$) and elevation 
($\theta_{c,s}^{\text{RIS}}$) arrival angles (with respect to the RIS broadside) and carefully calculated for our system due to the fixed orientation of the RIS. 
Here, the number of clusters, number of sub-rays per cluster, and the locations of the clusters can be determined for a given environment and frequency.

For the attenuation of the $(c,s)$th path, we adopt the 5G path loss model (the
close-in free space reference distance model with frequency-dependent path loss exponent, in dB), which is applicable to various environments including Urban Microcellular (UMi) and Indoor Hotspot (InH) \cite{5G_Channel}.
The LOS component of $\mathbf{h}$ is calculated by
\begin{equation}\label{eq:hLOS}
\mathbf{h}_{\text{LOS}}= I_{\mathbf{h}}(d_{\text{T-RIS}}) \sqrt{G_e(\theta_{\text{LOS}}^{\text{RIS}}) L_{\text{LOS}}^{\text{T-RIS}}}  e^{j\eta} \mathbf{a}(\phi_{\text{LOS}}^{\text{RIS}},\theta_{\text{LOS}}^{\text{RIS}})
\end{equation}
where $L_{\text{LOS}}^{\text{T-RIS}}$ is the attenuation of the LOS link, $G_e(\theta_{\text{LOS}}^{\text{RIS}})$ is the RIS element gain in the LOS direction, $\mathbf{a}(\phi_{\text{LOS}}^{\text{RIS}},\theta_{\text{LOS}}^{\text{RIS}})$ is the array response of the RIS in the direction of the Tx, and $\eta \sim \mathcal{U} [0,2\pi]$. Here, $\mathcal{U} [a,b]$ is a random variable uniformly distributed in $[a,b]$, and $I_{\mathbf{h}}(d_{\text{T-RIS}})$ is a Bernoulli random variable taking values from the set $\left\lbrace 0,1 \right\rbrace $ and characterizes the existence of a LOS link for a Tx-RIS separation of $d_{\text{T-RIS}}$. It is again calculated according to the 5G model \cite{5G_Channel}.

For the calculation of LOS-dominated RIS-Rx channel $\mathbf{g}$ in an indoor environment, we re-calculate the RIS array response in the direction of the Rx by calculating azimuth and elevation departure angles $\phi^{\text{RIS}}_{\text{Rx}}$ and $\theta^{\text{RIS}}_{\text{Rx}}$ for the RIS from the coordinates of the RIS and the Rx. Finally,  the vector of LOS channel coefficients can be generated as
\begin{equation}\label{eq:20}
\mathbf{g}=\sqrt{G_e(\theta^{\text{RIS}}_{\text{Rx}}) L_{\text{LOS}}^{\text{RIS-R}}} e^{j\eta} \mathbf{a}(\phi^{\text{RIS}}_{\text{Rx}},\theta^{\text{RIS}}_{\text{Rx}}).
\end{equation}
where $G_e(\theta^{\text{RIS}}_{\text{Rx}})$ is the gain of RIS element in the direction of the Rx, $L_{\text{LOS}}^{\text{RIS-R}}$ is the attenuation of LOS RIS-Rx channel, $\eta \sim \mathcal{U} [0,2\pi]$ is the random phase term and $\mathbf{a}(\phi^{\text{RIS}}_{\text{Rx}},\theta^{\text{RIS}}_{\text{Rx}})$ is the RIS array response in the direction of the Rx.

For outdoor channel modeling, the major change will be in the channel between the RIS and the Rx, which might be subject to small-scale fading as well with a random number of unique clusters. For this case, the RIS-Rx channel can be expressed as
\begin{equation}\label{eq:outdoor}
\mathbf{g}=
\bar{\gamma} \sum\limits_{c=1}^{\bar{C}} \sum\limits_{s=1}^{\bar{S_c}} \bar{\beta}_{c,s} \sqrt{G_e(\theta_{c,s}^{\text{Rx}}) L_{c,s}^{\text{Rx}}} \,\, \mathbf{a} ( \phi_{c,s}^{\text{Rx}}, \theta_{c,s}^{\text{Rx}} ) + \mathbf{g}_{\text{LOS}}
\end{equation}
where, similar to \eqref{eq:13}, $\bar{\gamma}$ is a normalization term, $\bar{C}$ and $\bar{S_c}$ stand for number of clusters and sub-rays per cluster for the RIS-Rx link, $\bar{\beta}_{c,s}$ is the complex path gain, $ L_{c,s}^{\text{Rx}} $ is the path attenuation, $G_e(\theta_{c,s}^{\text{Rx}})$ is the RIS element radiation pattern in the direction of the $(c,s)$th scatterer, $\mathbf{a} ( \phi_{c,s}^{\text{Rx}}, \theta_{c,s}^{\text{Rx}} )$ is the array response vector of the RIS for the given azimuth and elevation angles, and $\mathbf{g}_{\text{LOS}}$ is the LOS component. 

The RIS-assisted channel has a double-scattering nature, as a result, the single-scattering link between the Tx and Rx has to be taken into account in the proposed channel model. Even if the RIS is placed near the Rx, the Tx-Rx channel is relatively stronger than the RIS-assisted path, and cannot be ignored in the channel model. 

For indoors, using single-input single-output (SISO) mmWave channel modeling, the channel between these two terminals can be easily obtained (by ignoring arrival and departure angles) as
\begin{equation}\label{eq:18}
h_{\text{SISO}}=
\gamma \sum\limits_{c=1}^{C} \sum\limits_{s=1}^{S_c} \beta_{c,s} e^{j\eta_e} \sqrt{L_{c,s}^{\text{SISO}}} + h_{\text{LOS}}
\end{equation}
where $\gamma$, $C$, $S_c$, and $\beta_{c,s}$ are as defined in \eqref{eq:13} and remain the same for the Tx-Rx channel under the assumption of shared clusters with the Tx-RIS channel, while $h_{\text{LOS}}$ is the LOS component. Here, $L_{c,s}^{\text{SISO}}$ stands for the path attenuation for the corresponding link and $\eta_e$ is the excess phase caused by different travel distances of Tx-RIS and Tx-Rx links over the same scatterers. 

For outdoor environments, we assume that the RIS and the Rx are not too close to ensure that they have independent clusters (small scale parameters) as in the 3GPP 3D channel model \cite{3GPP_5G}. Using SISO mmWave channel modeling, the Tx-Rx channel can be easily obtained as
\begin{equation}\label{eq:SISO}
h_{\text{SISO}}=
\tilde{\gamma} \sum\limits_{c=1}^{\tilde{C}} \sum\limits_{s=1}^{\tilde{S}_c} \tilde{\beta}_{c,s} \sqrt{L_{c,s}^{\text{SISO}}} + h_{\text{LOS}}
\end{equation}
where the number of clusters $\tilde{C}$, sub-rays per cluster $\tilde{S}_c$, complex path gain $\tilde{\beta}_{c,s}$, and path attenuation $L_{c,s}^{\text{SISO}}$ are determined as discussed earlier for the Tx-RIS path and $ \tilde{\gamma}$ is the normalization term.

\begin{figure}[!t]
	\begin{center}
		\includegraphics[width=0.55\columnwidth]{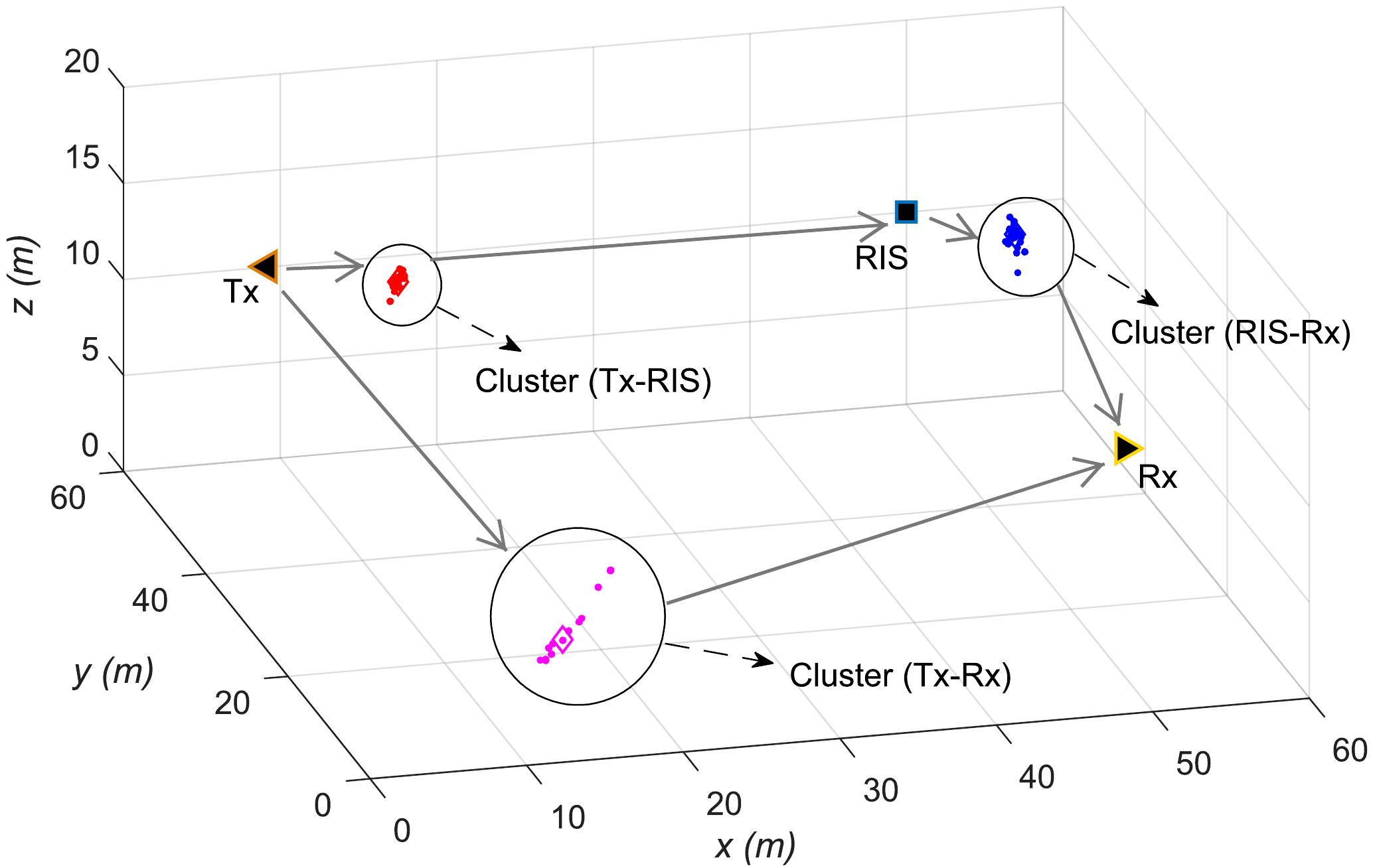}
		\vspace*{-0.3cm}\caption{The considered UMi Street Canyon outdoor environment with random number of clusters/scatterers and an RIS on the $xz$ plane. Source: Reprinted and adapted with permission from Basar, Yildirim and Kilinc \cite{SimRIS_TCOM}. Copyright 2021, IEEE.}
		\label{Fig3-3}
	\end{center}\vspace*{-0.5cm}
\end{figure}
In Fig. \ref{Fig3-3}, an exemplary 3D geometry for the UMi Street Canyon outdoor environment is illustrated for Scenario 1, where Tx is mounted at $20$ m height and Rx is a ground-level user. In this particular 3D geometry, each path has a single cluster with a different number of scatterers, while the number of clusters for each path can vary randomly in general.

\begin{figure*}
	\begin{center}
		\includegraphics[width=0.85\columnwidth]{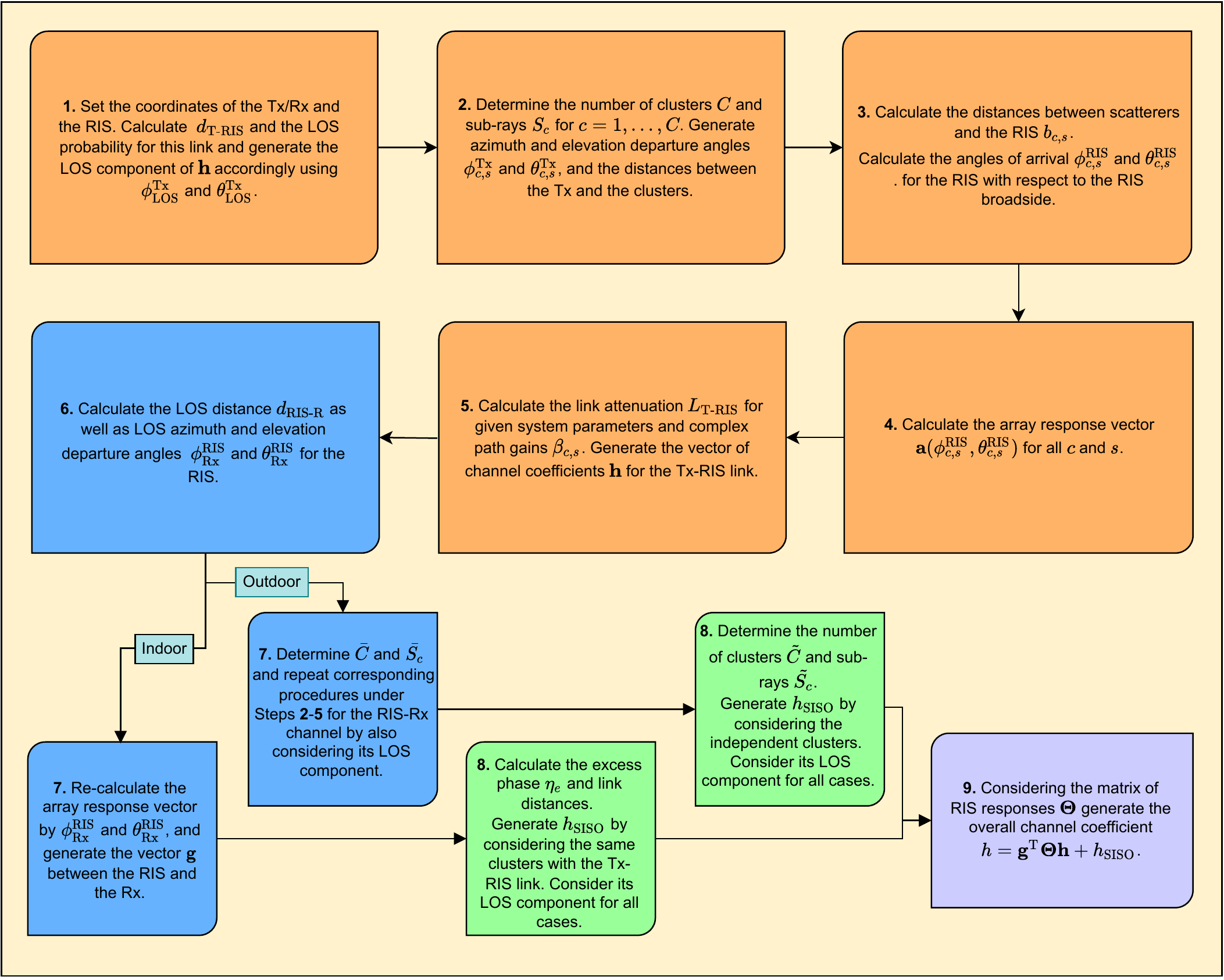}
		\vspace*{-0.3cm}\caption{Summary of the major steps of RIS-assisted physical channel modeling for indoor and outdoor environments.}\vspace*{-0.8cm}
		\label{fig3-1}
	\end{center}
\end{figure*}

The major steps of RIS-assisted physical channel modeling for indoor and outdoor environments can be summarized as in Fig. \ref{fig3-1}. In general, steps 1-5 focus on the generation of $\mathbf{h}$, while Steps 6-7 and Step 8 respectively deal with $\mathbf{g}$ and $h_{\text{SISO}}$.

Furthermore, in \cite{SimRIS_Mag} the proposed channel modeling strategy is extended to a MIMO system with $N_t$ transmit, and $N_r$ receive antennas operating in the presence of an RIS with $N$ reflecting elements. Consequently, the end-to-end channel matrix $\mathbf{C} \in\mathbb{C}^{N_r \times N_t}$ of an RIS-assisted MIMO system can be obtained as follows:
\begin{equation}
\mathbf{C}=\mathbf{G} \mathbf{\Phi} \mathbf{H} + \mathbf{D}. 
\end{equation}
In this model, $\mathbf{H}\in\mathbb{C}^{N \times N_t}$ is the matrix of channel coefficients between the Tx and the RIS, $\mathbf{G}\in\mathbb{C}^{N_r \times N}$ is the matrix of channel coefficients between the RIS and the Rx, and $\mathbf{D} \in\mathbb{C}^{N_r \times N_t}$ stands for the direct channel (not necessarily a LOS-dominated one and is likely to be blocked for mmWave bands due to obstacles in the environment) between the Tx and the Rx. The considered system configurations and general expressions of the channel matrices for the MIMO case are given in Table \ref{tab3-1}.

\begin{table}
\centering \small
\caption{System Configurations RIS-Assisted MIMO Channel Modeling.\label{tab3-1}}
\begin{threeparttable}
{%
\begin{tabular}{|p{0.13\textwidth}|p{0.75\textwidth}|}
\hline
\thead{Channel \\Matrices\tnote{*}}     &   \makecell{$\mathbf{H}= \gamma \sum\limits_{c=1}^{C} \sum\limits_{s=1}^{S_c} \beta_{c,s} \sqrt{G_e(\theta_{c,s}^{\text{T-RIS}})L_{c,s}^{\text{T-RIS}}} \,\, \mathbf{a} ( \phi_{c,s}^{\text{T-RIS}}, \theta_{c,s}^{\text{T-RIS}} ) \mathbf{a}^T( \phi_{c,s}^{\text{Tx}}, \theta_{c,s}^{\text{Tx}} ) + \mathbf{H}_{\text{LOS}}$\\
$\mathbf{G}=
\bar{\gamma} \sum\limits_{c=1}^{\bar{C}} \sum\limits_{s=1}^{\bar{S_c}} \bar{\beta}_{c,s} \sqrt{G_e(\theta_{c,s}^{\text{RIS-R}}) L_{c,s}^{\text{RIS-R}}} \,\, \mathbf{a} ( \phi_{c,s}^{\text{Rx}}, \theta_{c,s}^{\text{Rx}} ) \mathbf{a}^T( \phi_{c,s}^{\text{RIS-R}}, \theta_{c,s}^{\text{RIS-R}} ) + \mathbf{G}_{\text{LOS}}$\\
$\mathbf{D}=
\tilde{\gamma} \sum\limits_{c=1}^{\tilde{C}} \sum\limits_{s=1}^{\tilde{S}_c} \tilde{\beta}_{c,s} \sqrt{L_{c,s}^{\text{T-R}}}\mathbf{a} ( \phi_{c,s}^{\text{Rx'}}, \theta_{c,s}^{\text{Rx'}} ) \mathbf{a}^T( \phi_{c,s}^{\text{Tx'}}, \theta_{c,s}^{\text{Tx'}} ) + \mathbf{D}_{\text{LOS}}$} \\
\hline
\hline
\thead{Environments} & \makecell{Indoor: InH Indoor Office and Outdoor: UMi Street Canyon}\\
\hline
\thead{Frequencies} & \makecell{28 GHz and 73 GHz}    \\
\hline
\thead{Array Type} &\makecell{Uniform linear array (ULA) and uniform planar array (UPA)} \\
\hline
\end{tabular}}
 \begin{tablenotes}
            \item[*] $\gamma / \bar{\gamma}/ \tilde{\gamma}$: normalization factor \cite{SimRIS_Latincom}, $\beta_{c,s} / \bar{\beta}_{c,s}/\tilde{\beta}_{c,s}$: complex Gaussian distributed gain of the $(c,s)$th propagation path, $G_e(.)$: RIS element gain, $L^i_{c,s}$: attenuation of the $(c,s)$th propagation path \cite{3GPP_5G}, $\mathbf{a} \left( \phi^i_{c,s}, \theta^i_{c,s}\right)$: array response vectors for the considered azimuth ($\phi^i_{c,s}$) and elevation angles ($\theta^i_{c,s}$), $\mathbf{H}_{\text{LOS}}$/ $\mathbf{G}_{\text{LOS}}$/ $\mathbf{D}_{\text{LOS}}$: LOS component of sub-channels ($i$: indicator for the corresponding path/terminal as $i\in \lbrace \text{T-RIS},\text{RIS-R},\text{T-R},\text{Tx},\text{Rx}\rbrace$)
         
        \end{tablenotes}
     \end{threeparttable}
\end{table}


\section{Physical Channel Modeling for RIS-Empowered Systems at Sub-6 GHz Bands }
\begin{figure*}[!t]
	\begin{center}
		\includegraphics[width=1\columnwidth]{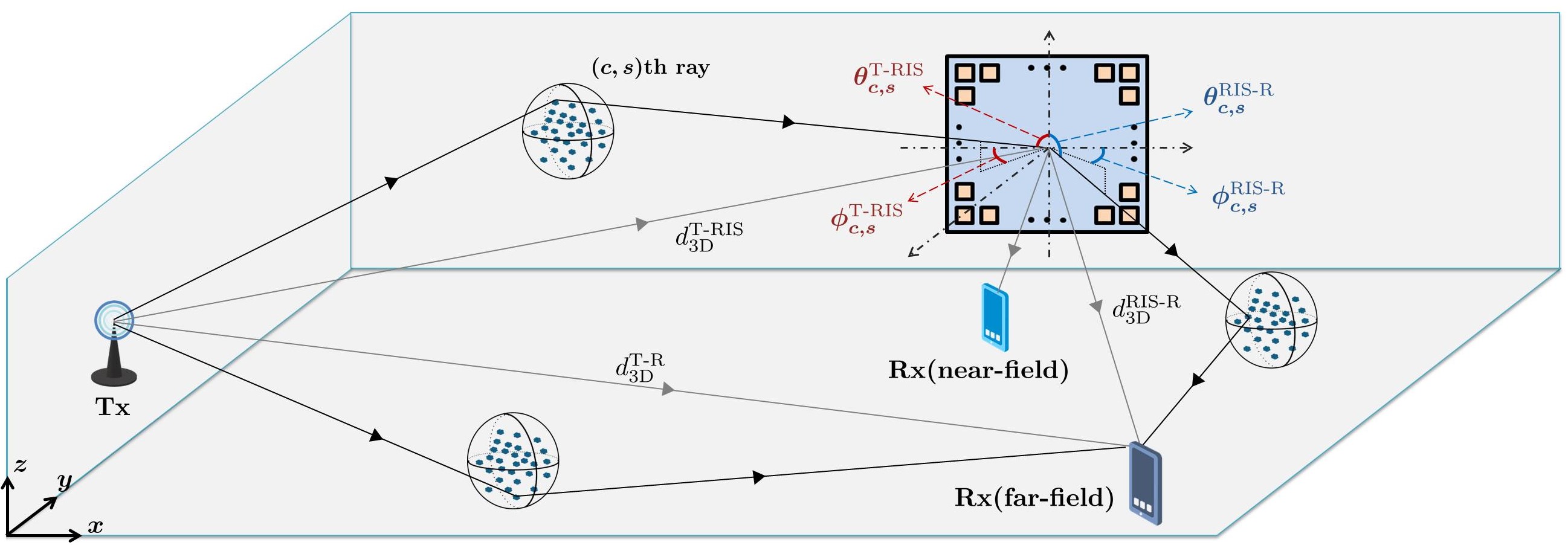}
		\vspace*{-0.3cm}\caption{Generic system model for an RIS-assisted network with $C$ number of clusters and $S$ number of rays. Source: Reprintedand adapted with permission from Kilinc, Yildirim and Basar \cite{Kilinc_2021}. Copyright 2021, IEEE.}\vspace*{-0.3cm}
		\label{fig:Sysmodel}
	\end{center} \vspace*{-0.4cm}
\end{figure*}

This section introduces a channel modeling strategy for RIS-assisted wireless networks in sub-6 GHz bands by investigating far-field and near-field behaviors in transmission. In addition to mmWave bands, RIS-assisted communication systems draw attention as an effective solution in sub-6 GHz frequencies, which are widely used in wireless communications. According to the technical specifications on sub-6 GHz bands \cite{3GPP_Sub6G,article}, an end-to-end channel model is derived for SISO systems employing an RIS by following the 3D channel modeling approach. The generic system model for the considered RIS-assisted wireless communication system is demonstrated in Fig. \ref{fig:Sysmodel}, where $d_\text{3D}^\text{T-RIS}$,$d_\text{3D}^\text{T-R}$ and $d_\text{3D}^\text{RIS-R}$ denotes the 3D distances between the Tx-RIS, Tx-Rx and RIS-Rx, respectively. It is assumed that the RIS with $N$ number of elements is located on the $xz$-plane. There exist $C$ clusters in the environment, each containing $S$ rays. The channels between the Tx-RIS, RIS-Rx and Tx-Rx are represented by $\mathbf{h}\in \mathbb{C}^{N\times 1}$, $\mathbf{g}\in \mathbb{C}^{N\times 1}$, and $h_\text{SISO}\in \mathbb{C}^{1\times 1}$, respectively, where $N$ is the number of RIS elements. It is assumed that the Tx and Rx are equipped with unity gain isotropic antennas. The positions of the Tx, Rx and RIS are given in the cartesian coordinate system as $\textbf{r}^\text{Tx}=(x^{\text{Tx}},y^{\text{Tx}},z^{\text{Tx}})$, $\textbf{r}^\text{Rx}=(x^{\text{Rx}},y^{\text{Rx}},z^{\text{Rx}})$ and $\textbf{r}^\text{RIS}=(x^{\text{RIS}},y^{\text{RIS}},z^{\text{RIS}})$, respectively. 

\begin{table}
\centering \footnotesize
\caption{System Parameters of Physical RIS-Assisted Channel Modeling for sub-6 GHz.\label{tab4-1}}{%
\begin{tabular}{|c|l|}
\hline
$P_c$ & Power of the $n$th cluster \\
\hline
$P_L$ &  Path loss component \\
\hline
$G_e(\theta_{c,s}^{\text{T-RIS}})$ & Radiation pattern of an RIS element in the direction of $(c,s)$th path\\
\hline
$\Phi_{c,s}$& Random initial phase \\
\hline
$\mathbf{a}(\theta_{c,s}^{\text{T-RIS}},\phi_{c,s}^{\text{T-RIS}})$& Array response vector of the RIS in the direcion of the Tx\\
\hline
$\theta_{c,s}^{\text{T-RIS}}$& Zenith angle of arrival (ZoA) \\
\hline
$\phi_{c,s}^{\text{T-RIS}}$& Azimuth angle of arrival (AoA) of the RIS for the $(c,s)$th path \\
\hline
$G_e(\theta_{c,s}^{\text{RIS-R}})$ & RIS element radiation pattern in the direction of $(c,s)$th path \\
\hline
$\phi_{c,s}^{\text{RIS-R}}$ & Azimuth of departure (AoD) and from the RIS \\
\hline
$\theta_{c,s}^{\text{RIS-R}}$ &  Zenith of departure (ZoD) angles from the RIS \\
\hline
 $\mathbf{a}(\theta_{c,s}^{\text{RIS-R}},\phi_{c,s}^{\text{RIS-R}})$ & Array response vector of the RIS in the direction of the Rx\\
\hline
\end{tabular}}{}
\end{table}

\begin{figure*}[!t]
	\begin{center}
		\includegraphics[width=0.9\columnwidth]{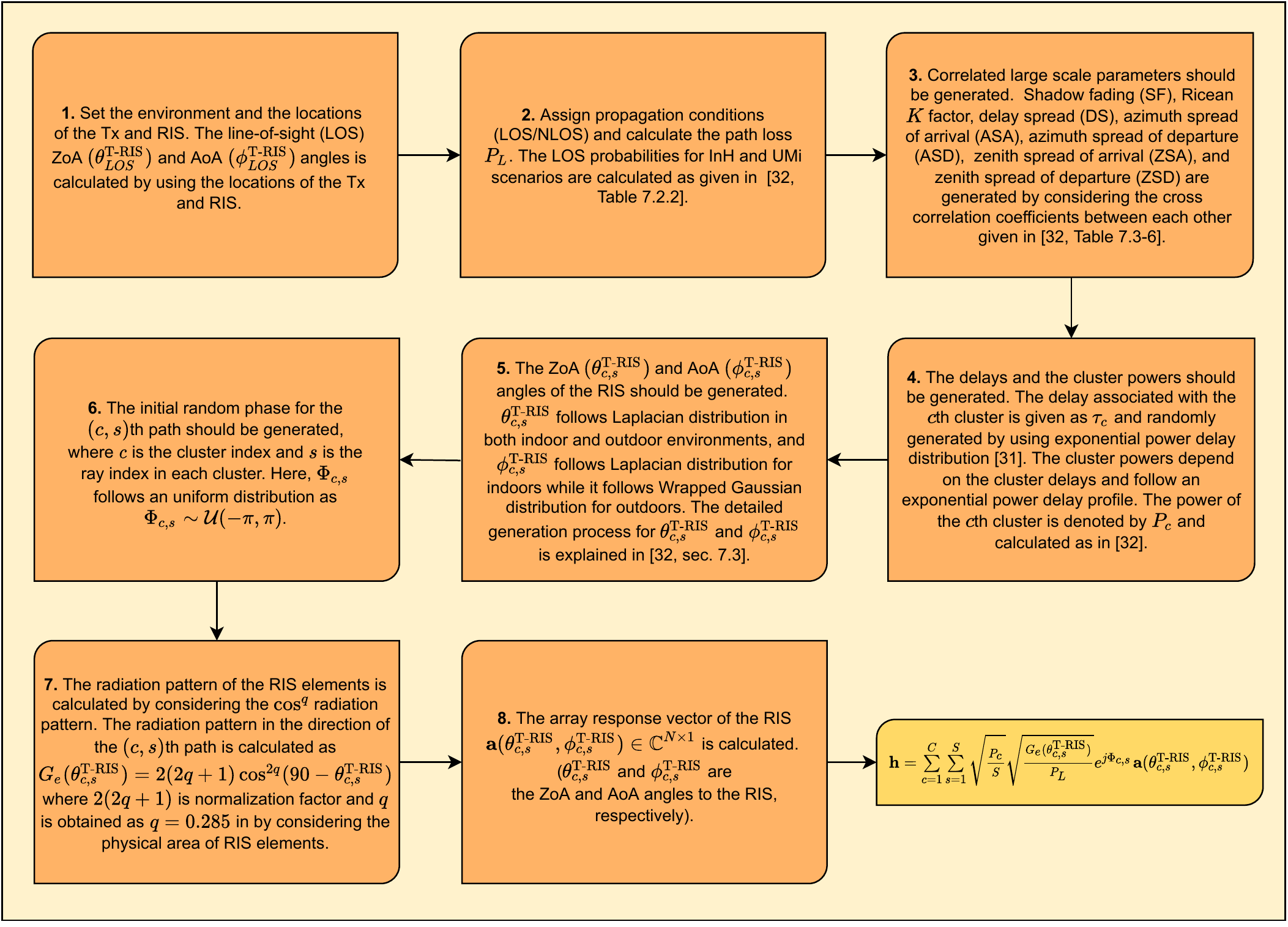}
		\vspace*{-0.3cm}\caption{The major generation steps of the Tx-RIS channel in RIS-assisted physical channel modeling.}\vspace*{-0.8cm}
		\label{fig4-2}
	\end{center}
\end{figure*}

The definitions of the system parameters are given in Table \ref{tab4-1} for the RIS-assisted channels. The major generation steps of the Tx-RIS channel in RIS-assisted physical channel modeling are also summarized in Fig. \ref{fig4-2}. Interested readers are referred to \cite{Kilinc_2021} for the details channel generation procedure and statistical background of the considered channel parameters. By considering the certain system parameters and following the channel generation steps in Fig. \ref{fig4-2},  the  channel  between  the  Tx  and  RIS is expressed as
\begin{equation}\label{eq:h}
    \mathbf{h}=\sum\limits_{c=1}^{C}\sum\limits_{s=1}^{S}\sqrt{\frac{P_c}{S}\vphantom{\frac{G_e(\theta_{c,s}^{\text{T-RIS}})}{P_L}}}\sqrt{\frac{G_e(\theta_{c,s}^{\text{T-RIS}})}{P_L}}e^{j\Phi_{c,s}}\mathbf{a}(\theta_{c,s}^{\text{T-RIS}},\phi_{c,s}^{\text{T-RIS}}).
\end{equation}

Furthermore, the direct link between the Tx and Rx can be obtained by following the Steps 1, 2, 3, 4 and 6 in the channel generation procedure described in Fig. \ref{fig4-2}. Therefore, the direct link between the Tx and Rx is also given by
\begin{align}
    h_\text{SISO}=\sum\limits_{c=1}^{C}\sum\limits_{s=1}^{S}\sqrt{\frac{P_c}{S}}\sqrt{\frac{1}{P_L}}e^{j\Phi_{c,s}}
\end{align}
Here, in all steps, the Tx and Rx positions should be taken into account instead of Tx and RIS positions, respectively. Moreover,  AoA and ZoA angles calculations are not required in the generation of $h_\text {SISO}$, since the SISO (direct) link is established.

For the RIS-Rx channel, two different channel scenarios are separately considered between the RIS and Rx. If the size of the RIS is large enough, the far-field boundary between the RIS and Rx will be extremely large. Accordingly, the near-field cases are also included when the RIS is placed close proximity of the Rx, while far-field channels are obtained by following the similar procedure in Fig. \ref{fig4-2}. Under the far-field conditions, the channel between the RIS and the Rx is expressed as 
\begin{align} 
    \mathbf{g}=\sum\limits_{c=1}^{C}\sum\limits_{s=1}^{S}\sqrt{\frac{P_c}{S}\vphantom{\frac{G_e(\theta_{c,s}^{\text{T-RIS}})}{P_L}}}\sqrt{\frac{G_e(\theta_{c,s}^{\text{RIS-R}})}{P_L}}e^{j\Phi_{c,s}}\mathbf{a}(\theta_{c,s}^{\text{RIS-R}},\phi_{c,s}^{\text{RIS-R}}).
\end{align}
The generation steps for the far-field RIS-Rx channel  are similar to the channel generation steps of the Tx-RIS link. Differently, the RIS and Rx positions should be considered instead of Tx and RIS positions, respectively. In Steps 5, 7 and 8, the angles $\theta_{c,s}^{\text{RIS-R}}$ and $\phi_{c,s}^{\text{RIS-R}}$ should be taken into account instead of $\theta_{c,s}^{\text{T-RIS}}$ and $\phi_{c,s}^{\text{T-RIS}}$, and the distributions of them will be same with $\theta_{c,s}^{\text{T-RIS}}$ and $\phi_{c,s}^{\text{T-RIS}}$, respectively.

The near-field channel between the RIS and Rx is denoted by 
\begin{equation}
    \textbf{g}=[g_1,g_2\cdots,g_N]^\mathrm{T}
\end{equation} 
where $g_n$ represents the channel coefficient from the $n$th RIS element to the Rx, and can be expressed in terms of its magnitude and phase as $g_n=|g_n|e^{-j\gamma}$ for $n= 1,2,\dots,N$.
By considering the RIS geometry and the RIS element locations in \cite{Kilinc_2021}, the near field channel gain from the $n$th RIS element is approximated as  \cite{Emilpower_2020}
\begin{align}\label{eq:nf}
    |g_n|^2\approx\frac{1}{4\pi}\sum\limits_{x\in\mathbb{X}}\sum\limits_{z\in\mathbb{Z}} \left(\frac{\frac{xz}{y^2}}{3\left( \frac{z^2}{y^2}+1\right)\sqrt{\frac{x^2}{z^2}+\frac{z^2}{y^2}+1}}\right.
    \left.+\frac{2}{3}\tan^{-1}\left( \frac{\frac{xz}{y^2}}{\sqrt{\frac{x^2}{y^2}+\frac{z^2}{y^2}+1}}\right)\right)
\end{align}
where $\mathbb{X}=\left\lbrace d/2+x^n-x^\text{Rx},d/2+x^\text{Rx}-x^n\right\rbrace$, $\mathbb{Z}=\lbrace d/2+z^n-z^\text{Rx},d/2+z^\text{Rx}-z^n\rbrace$, and $y=|y^n-y^\text{Rx}|$.
Moreover, the phase of $g_n$ can be calculated as follows:
\begin{align}
    \gamma=2\pi\hspace{-0.3 cm}\mod{\left(\frac{\|\textbf{r}^n-\textbf{r}^\text{Rx}\|}{\lambda},1\right)}.
\end{align} 



\section{SimRIS Channel Simulator}
This section introduces the open-source, user-friendly, and widely applicable \textit{SimRIS Channel Simulator v2.0} \cite{SimRIS_Latincom,SimRIS_Mag}. \textit{SimRIS Channel Simulator}  aims to open a new line of research for physical channel modeling of RIS-empowered networks by including the LOS probabilities between terminals, array responses of RISs and Tx/Rx units, RIS element gains, realistic path loss and shadowing models, and environmental characteristics in different propagating environments and operating frequencies. More specifically, Indoor Hotspot (InH) -  Indoor Office and Urban Microcellar (UMi) - Street Canyon environments are considered for popular mmWave operating frequencies of 28 GHz and 73 GHz from the 5G channel model \cite{3GPP_5G}. Compared to its earlier version (\textit{SimRIS Channel Simulator v1.0}) with only SISO Tx/Rx terminals, its v2.0 considers MIMO terminals with  different types of arrays \cite{SimRIS_Mag}. The simulator also supports terminals in the far-field of the RIS only, which is a reasonable assumption for mmWaves with shorter wavelengths and smaller RIS sizes, while the near-field RIS models will be easily included by extending the existing modeling strategy in its future releases.

\begin{figure}[!t]
	\begin{center}
		\includegraphics[width=0.75\columnwidth]{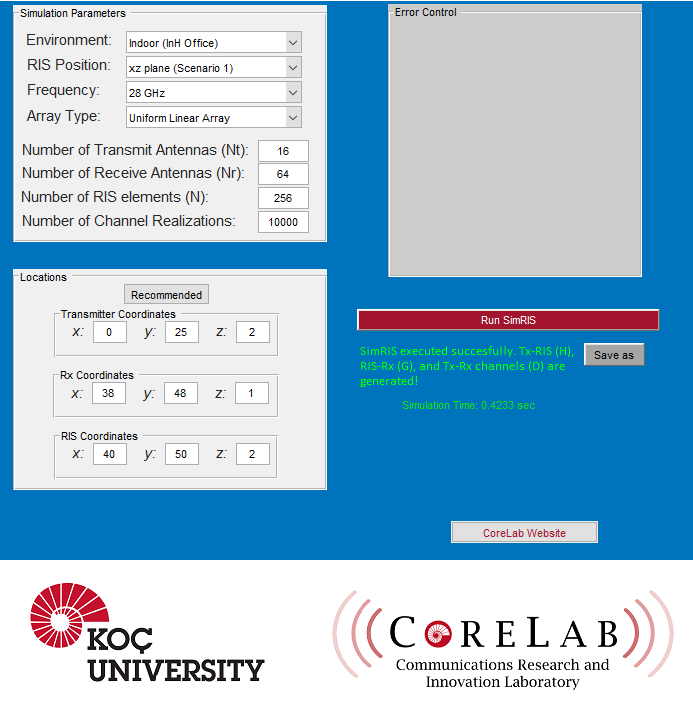}
		\vspace*{-0.1cm}\caption{ Graphical user interface (GUI) of the SimRIS Channel Simulator v2.0.}\vspace*{-0.7cm}
		\label{fig5-1}
	\end{center}
\end{figure}

The considered generic 3D geometry is given in Fig. \ref{fig3-2} for the representation of physical channel characteristics. In this setup, the RIS is mounted on the $xz$-plane while a generalization is possible in the simulator. The proposed model considers various indoor and outdoor wireless propagation environments in terms of physical aspects of mmWave frequencies while numerous practical 5G channel model issues are adopted to our channel model \cite{3GPP_5G}. For a considered operating frequency and environment, the number of clusters ($C$), number of sub-rays per cluster ($S_c$), and the positions of the clusters can be determined by following the detailed steps and procedures in \cite{SimRIS_TCOM}. More specifically, according to the 5G channel model, the number of clusters and scatterers are determined using the Poisson and uniform distributions with certain parameters, respectively. Although the clusters between Tx-RIS and Tx-Rx can be modeled independently, it can be assumed that the Rx and the RIS might share the same clusters when the Rx is located relatively closer to the RIS. 
 Due to the fixed orientation of the Tx and the RIS, the array response vectors of the Tx and the RIS are easily calculated for given azimuth and elevation departure/arrival angles. However, it is worth noting that if azimuth and elevation angles are generated randomly for the Tx, due to fixed orientation of the RIS, they will not be random anymore at the RIS and should be calculated from the 3D geometry using trigonometric identities. Nevertheless, the array response vector of the Rx can be calculated with randomly distributed azimuth and elevation angles of arrival due to the random orientation.

Using the SimRIS Channel Simulator, the wireless channels of RIS-aided communication systems can be generated with tunable operating frequencies, number of RIS elements, number of transmit/receive antennas, Tx/Rx array types, terminal locations, and environments. As discussed earlier, in SimRIS Channel Simulator v2.0, MIMO-aided Tx and Rx terminals are incorporated into its earlier version, and the array response vectors and receiver orientation are reconstructed by considering this MIMO system model. This new version also offers two different types of antenna array configurations: Uniform linear array (ULA) and uniform planar array (UPA), and the corresponding array response vectors are calculated according to the selected antenna array type. The graphical user interface (GUI) of our SimRIS Channel Simulator v2.0 is given in Fig. \ref{fig5-1}. In this GUI, the selected scenario specifies the RIS position for both indoor and outdoor environments as $xz$-plane (Scenario 1) and $yz$-plane (Scenario 2). Considering the 3D geometry illustrated in Fig. \ref{fig3-2}, Tx, Rx, and RIS positions can be manually entered into the SimRIS Channel Simulator. Furthermore,  $N_t$ (the number of Tx antennas), $N_r$ (the number of Rx antennas), $N$ (the number of RIS elements), and the number of channel realizations are also user-selectable input parameters and these options offer a flexible and versatile channel modeling opportunity to the users.  Considering these input parameters, this simulator produces $\mathbf{H}$, $\mathbf{G}$, and $\mathbf{D}$ channel matrices by conducting Monte Carlo simulations for the specified number of realizations under $28$ and $73$ GHz mmWave frequencies. The general expressions of these channel matrices are given in Table 1 for the interested readers. Here, the double summation terms stem from the random number of clusters and scatterers and the LOS components might be equal to zero with a certain probability for increasing distances. 

Consequently, the open-source SimRIS Channel Simulator package considers a narrowband channel model for RIS-empowered communication systems for both indoor and outdoor environments and it takes into account various physical characteristics of the wireless propagation environment.  The open-source nature of our simulator, which is written in the MATLAB programming environment, encourages all researchers to use and contribute to the development of its future versions by exploring the interesting use-cases of the RIS in the transmission.
\section{Performance Analysis Using SimRIS Channel Simulator}

\begin{table}
\centering \small
\caption{System Configurations and Simulation Parameters.\label{tab6-1}}{%
\begin{tabular}{|c||c|c|c|}
\hline
  \textbf{Setup} &  $\mathbf{1}^\textbf{st}$ & $\mathbf{2}^\textbf{nd}$ & $\mathbf{3}^\textbf{rd}$ \\
\hline
\hline
\textbf{Frequency}      &   28 GHz                                          & 28 GHz & 28 GHz\\
\hline
\textbf{Environments}   & {InH Indoor Office}                               & InH Indoor Office&InH Indoor Office\\
\hline
\textbf{$N$}            & 256                                               & Varying& 64\\
\hline
\textbf{$N_t=N_r$}      & 1                                                 & Varying & 4\\
\hline
\textbf{Array Type}     & -                                                 & UPA & ULA\\
\hline
\textbf{Tx Position}    & [0,25,2]                                          & [0,25,2]& [0,25,2]  \\
\hline
\textbf{Rx Position}    & Varying                                           & [45,45,1] & Varying \\
\hline
\textbf{RIS Positions}   &\makecell{RIS 1: [40,50,2] \\ RIS 2: [60,40,2.5]}  & [40,50,2]  &\makecell{RIS 1: [40,46,2] \\ RIS 2: [62,30,2]}\\
\hline
\end{tabular}}
\end{table}
This section provides numerical results that are conducted via \textit{SimRIS Channel Simulator} MATLAB package for the detailed evaluation of how RISs can be effectively used in future wireless networks to enrich and improve the existing communication systems. The  system configurations and computer simulation parameters of the considered setups are given in Table \ref{tab6-1}, which will be used for the numerical results in the following. All computer simulations in this section are conducted in an InH Indoor Office environment at an operating frequency of $28$ GHz, while the considered system parameters are also valid for $73$ GHz and outdoor environments. The positions of the terminals are given in the 3D Cartesian coordinate system, and the noise and transmit powers are assumed to be -$100$ dBm and $30$ dBm, respectively.

\begin{figure}[!t]
	\begin{center}
		\includegraphics[width=0.8\columnwidth]{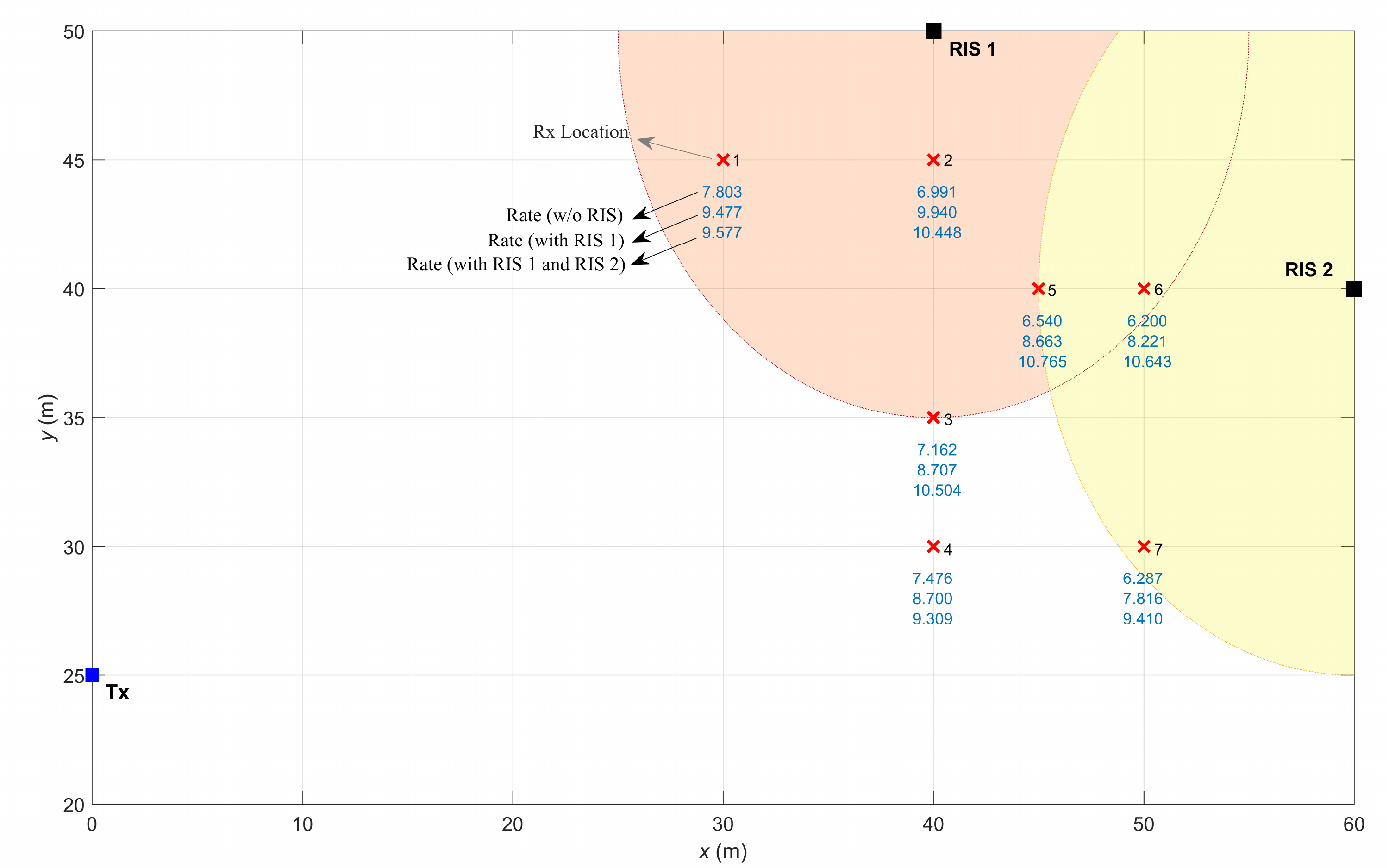}
		\vspace*{-0.1cm}\caption{Top view of the considered transmission scenario with 7 reference points along with the achievable rate values.}\vspace*{-0.7cm}
		\label{fig6-1}
	\end{center}
\end{figure}

In Fig. \ref{fig6-1}, the achievable rate performance is investigated for 7 different positions of the Rx in an InH Indoor Office environment under SISO case. Here, $(x^{\text{Rx}},y^{\text{Rx}})$ coordinates of the test points are marked on Fig. \ref{fig6-1}, while $z^{\text{Rx}}$ is fixed to $1$ m for all points. For this analysis, the first setup parameters in Table \ref{tab6-1} are considered. In order to observe the effect of RIS in transmission, achievable rates of these 7 reference positions of the Rx are calculated for three cases: RISs are not used, one RIS is used, and two RISs are used. 
In the case of only RIS 1 being operated, an average increase of $26.91\%$ is achieved in the achievable rate of the reference points, while an increase of $45.72\%$ is achieved when RIS 1 and RIS 2 are jointly activated. Furthermore, when RISs are deactivated, reference points 1, 3, and 4 will have a higher achievable rate since they are close to the Tx compared to other points, while all points will have approximately similar achievable rate performance when RIS 1 and RIS 2 are activated. As observed from Fig. \ref{fig6-1}, we concluded that a significant improvement is obtained in achievable rate with RISs, particularly for the test points closer to the RIS. 

\begin{figure}[h]
	\begin{center}
		\includegraphics[width=0.60\columnwidth]{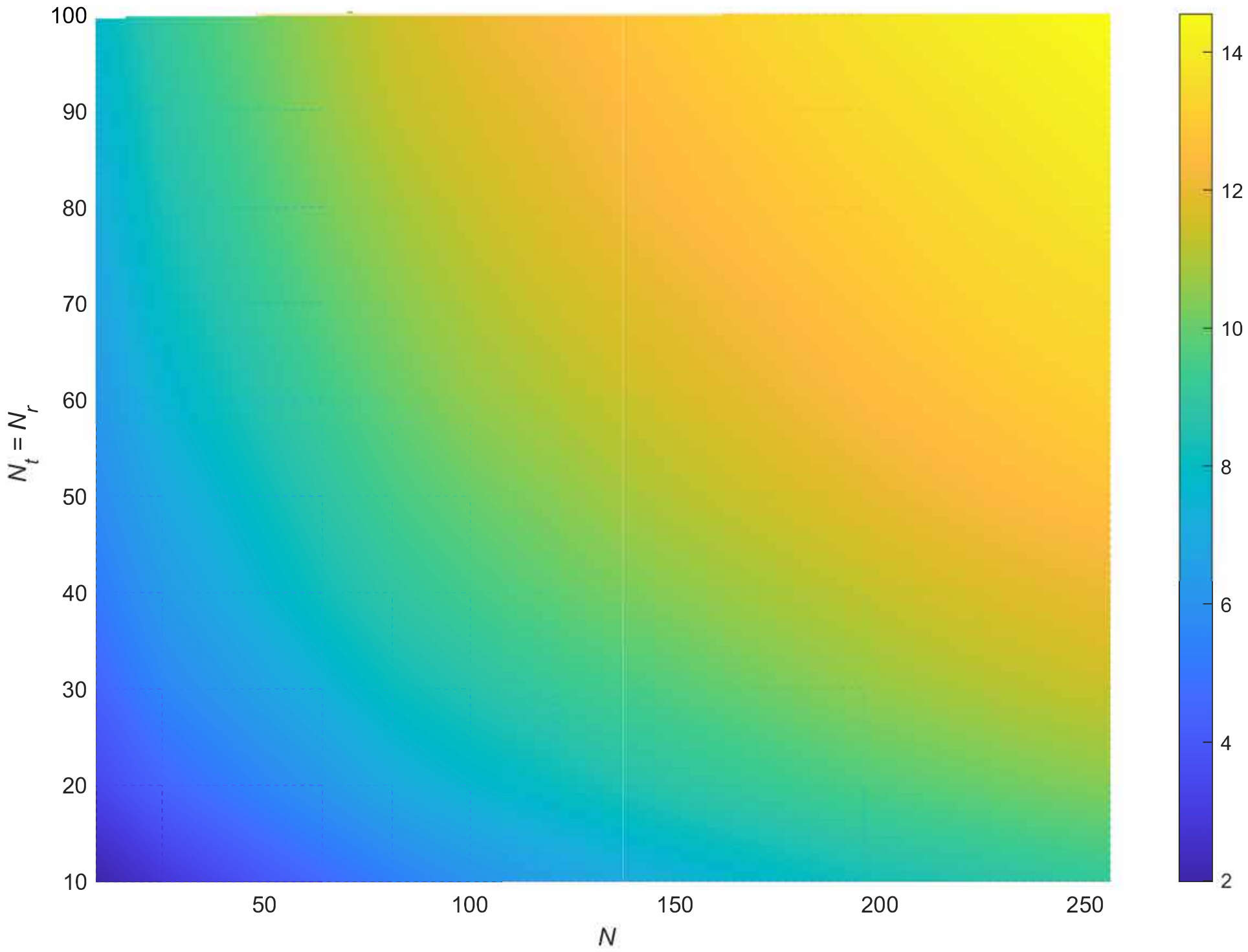}
		\vspace*{-0.1cm}\caption{ Achievable rate analysis in the presence of an RIS for changing $N_t$/$N_r$ and $N$.}\vspace*{-0.7cm}
		\label{fig6-2}
	\end{center}
\end{figure}

In order to get a more precise understanding of the impacts of the reflecting elements and number of Tx/Rx antennas on the system performance, the combined effect of these two parameters on the achievable rate is analyzed in Fig. \ref{fig6-2}. The second setup parameters in Table \ref{tab6-1} are considered for this analysis, and pseudoinverse (pinv)-based algorithm \cite{Pinv_2019} is used for adapting the phase shifts of an RIS-assisted MIMO transmission system. As observed from the Fig. \ref{fig6-2}, doubling $N_t$ and $N_r$ values for $N=2$5 provides approximately $2.09$ bit/sec/Hz increase in achievable rate, while doubling $N$ values for $N_t=N_r=20$ provides a roughly $2.01$ bit/sec/Hz increase in achievable rate. In order to meet the increasing data demand in next generation wireless networks, it is foreseen to use a large number of transmit and receive antennas. Although, using a large number of antennas enhance the achievable rate performances, the cost of signal processing and hardware in will be notably ascended. Using RISs in the transmission also provides enhanced achievable rate, while alleviating the cost of physical implementation and signal processing. From the obtained results in \ref{fig6-2}, it can be said that the demand that will emerge in future wireless networks can be satisfied effectively by increasing the $N$, since doubling the number of reflectors is much less costly than doubling the number of Tx/Rx antennas. 

\begin{figure}[!t]
	\begin{center}
		\includegraphics[width=0.8\columnwidth]{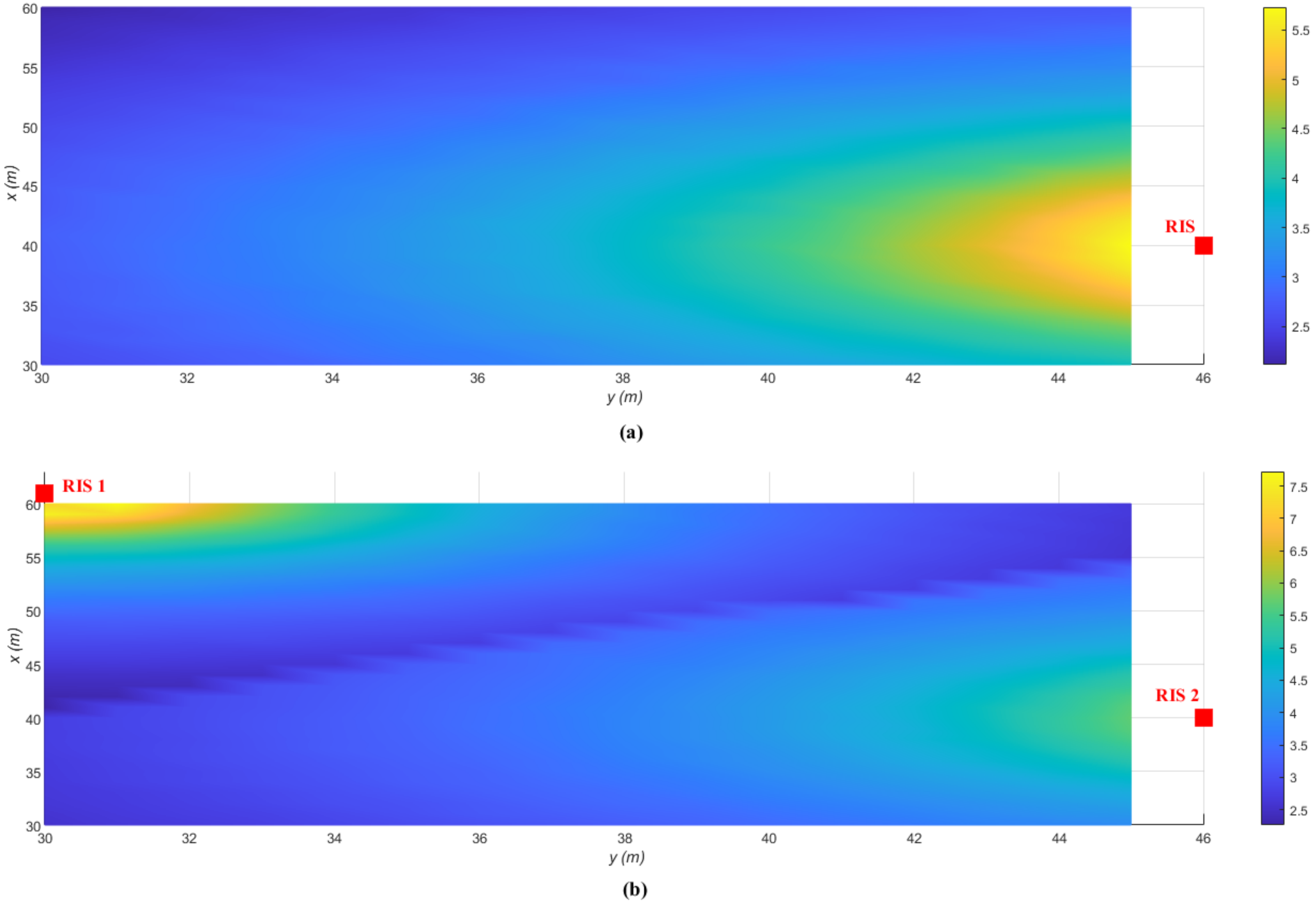}
		\vspace*{-0.1cm}\caption{ Achievable rate analysis of MIMO communication system with varying Rx positions in the presence of (b) A single RIS and (c) Two RISs. Note: The red squares indicate the positions of the RISs on the $x$ and $y$-axes.}\vspace*{-0.7cm}
		\label{fig6-3}
	\end{center}
\end{figure}

By considering the third setup parameters in Table \ref{tab6-1}, the effect of the changing Rx positions on the $xy$-plane to its achievable rate performance is analyzed in Fig.  \ref{fig6-3} in the presence of a single and two RISs under MIMO transmission scenario. In Fig. \ref{fig6-3}(a), the indoor coverage extension is provided by using an RIS when the direct link between the Tx and Rx are blocked due to the obstacles in the environment. Therefore, it can be obtained from Fig. \ref{fig6-3}(a) that particularly in regions where Rx is close to RIS, a significant increase in achievable rate is achieved, guaranteeing reliable communication even if there is no direct link between the Tx and Rx. If a larger area is desired to be covered, the idea of using more than one RIS for transmission may stand out in the system design. Within this context, the achievable rate performance for  the varying Rx locations is observed in Fig. \ref{fig6-3}(b). Here, it is aimed to enhance the received signal quality by modifying the phase shifts of the RISs, which are closer to the Rx. Particularly, employing two RISs create a handover capability for the Rx to increase its signal power. As obtained from Fig. \ref{fig6-3}(b), a considerable increase in the achievable rate is obtained when the Rx is placed to the proximity of any of the RISs. From the given results,  the use of RISs in transmission boosts signal quality even in dead zones or cell edges and provides a coverage extension by providing low-cost and energy-efficient solutions by alleviating a large number of antenna requirements in future wireless networks.

\section{Summary}

In this work,  we mainly aim to present a general framework on channel modeling strategies for the RIS-empowered communications systems considering the state-of-the-art channel and propagation modeling efforts in the literature. Another objective of this work is to draw attention to open-source and standard-compliant physical channel modeling efforts to provide comprehensive insights regarding the practical use-cases of  RISs in future wireless networks. Within this context, a framework on the cluster-based statistical channel model is summarized for both sub-6 GHz and mmWave bands. Moreover, an open-source \textit{SimRIS Channel Simulator}  package that considers a narrowband channel model for RIS-empowered communication systems is presented. Finally, using \textit{SimRIS Channel Simulator},  performances of the RIS-assisted communication system are evaluated, and efficient use-cases of RISs for next-generation wireless systems are investigated.

\bibliographystyle{IEEEtran}
\bibliography{bib_2022}

\begin{thebibliography}{10}
\providecommand{\url}[1]{#1}
\csname url@samestyle\endcsname
\providecommand{\newblock}{\relax}
\providecommand{\bibinfo}[2]{#2}
\providecommand{\BIBentrySTDinterwordspacing}{\spaceskip=0pt\relax}
\providecommand{\BIBentryALTinterwordstretchfactor}{4}
\providecommand{\BIBentryALTinterwordspacing}{\spaceskip=\fontdimen2\font plus
\BIBentryALTinterwordstretchfactor\fontdimen3\font minus
  \fontdimen4\font\relax}
\providecommand{\BIBforeignlanguage}[2]{{%
\expandafter\ifx\csname l@#1\endcsname\relax
\typeout{** WARNING: IEEEtran.bst: No hyphenation pattern has been}%
\typeout{** loaded for the language `#1'. Using the pattern for}%
\typeout{** the default language instead.}%
\else
\language=\csname l@#1\endcsname
\fi
#2}}
\providecommand{\BIBdecl}{\relax}
\BIBdecl

\bibitem{Samsung_2020}
\BIBentryALTinterwordspacing
``Samsung's {6G} white paper lays out the company's vision for the next
  generation of communications technology,'' July 2020. [Online]. Available:
  \url{https://research.samsung.com/next-generation-communications}
\BIBentrySTDinterwordspacing

\bibitem{Rajatheva_6G}
\BIBentryALTinterwordspacing
N.~{Rajatheva \it{et al.}}, ``White paper on broadband connectivity in 6{G},''
  Apr. 2020. [Online]. Available: \url{http://arxiv.org/abs/2004.14247}
\BIBentrySTDinterwordspacing

\bibitem{SimRIS_Mag}
E.~Basar and I.~Yildirim, ``Reconfigurable intelligent surfaces for future
  wireless networks: {A} channel modeling perspective,'' \emph{IEEE Wireless
  Commun.}, vol.~28, no.~3, pp. 108--114, 2021.

\bibitem{Basar_Access_2019}
E.~{Basar \it{et al.}}, ``Wireless communications through reconfigurable
  intelligent surfaces,'' \emph{IEEE Access}, pp. 116\,753--116\,773, Sep.
  2019.

\bibitem{Wu_Tutorial}
Q.~{Wu \it{et al.}}, ``Intelligent reflecting surface aided wireless
  communications: {A} tutorial,'' \emph{IEEE Trans. Commun.}, vol.~69, no.~5,
  pp. 3313--3351, 2021.

\bibitem{Basar_2019_LIS_2}
\BIBentryALTinterwordspacing
E.~Basar, ``Reconfigurable intelligent surface-based index modulation: {A} new
  beyond {MIMO} paradigm for 6{G},'' Apr. 2019. [Online]. Available:
  \url{arXiv:1904.06704}
\BIBentrySTDinterwordspacing

\bibitem{Yildirim_hybrid}
I.~Yildirim, F.~Kilinc, E.~Basar, and G.~C. Alexandropoulos, ``Hybrid
  {RIS}-empowered reflection and decode-and-forward relaying for coverage
  extension,'' \emph{IEEE Commun. Lett.}, vol.~25, no.~5, pp. 1692--1696, 2021.

\bibitem{Yildirim_multiRIS}
I.~Yildirim, A.~Uyrus, and E.~Basar, ``Modeling and analysis of reconfigurable
  intelligent surfaces for indoor and outdoor applications in future wireless
  networks,'' \emph{IEEE Trans. Commun.}, vol.~69, no.~2, pp. 1290--1301, 2021.

\bibitem{Arslan_2021}
\BIBentryALTinterwordspacing
E.~Arslan, I.~Yildirim, F.~Kilinc, and E.~Basar, ``Over-the-air equalization
  with reconfigurable intelligent surfaces,'' June 2021. [Online]. Available:
  \url{https://arxiv.org/abs/2106.07996}
\BIBentrySTDinterwordspacing

\bibitem{Renzo_Analy2020}
M.~{Di Renzo \it{et al.}}, ``Analytical modeling of the path-loss for
  reconfigurable intelligent surfaces – {A}nomalous mirror or scatterer?'' in
  \emph{IEEE Int. Workshop Signal Process. Adv. Wireless Commun. (SPAWC)},
  2020, pp. 1--5.

\bibitem{Garcia_2019}
J.~B. Garcia, A.~Sibille, and M.~Kamoun, ``Reconfigurable intelligent surfaces:
  Bridging the gap between scattering and reflection,'' \emph{IEEE J. Sel.
  Areas Commun}, vol.~38, no.~11, pp. 2538--2547, Nov. 2020.

\bibitem{Ellingson}
\BIBentryALTinterwordspacing
S.~{Ellingson}, ``Path loss in reconfigurable intelligent surface-enabled
  channels,'' Nov. 2021. [Online]. Available:
  \url{http://arxiv.org/abs/1912.06759}
\BIBentrySTDinterwordspacing

\bibitem{Ozdogan_2020}
O.~{Ozdogan}, E.~{Bjornson}, and E.~G. {Larsson}, ``Intelligent reflecting
  surfaces: {P}hysics, propagation, and pathloss modeling,'' \emph{IEEE
  Wireless Commun. Lett.}, vol.~9, no.~5, pp. 581--585, May 2020.

\bibitem{Tang_2020}
W.~{Tang \it{et al.}}, ``Wireless communications with reconfigurable
  intelligent surface: {P}ath loss modeling and experimental measurement,''
  \emph{IEEE Trans. Wireless Commun.}, vol.~20, no.~1, pp. 421--439, 2021.

\bibitem{Danufane_2021}
F.~H. Danufane, M.~D. Renzo, J.~de~Rosny, and S.~Tretyakov, ``On the path-loss
  of reconfigurable intelligent surfaces: {A}n approach based on green’s
  theorem applied to vector fields,'' \emph{IEEE Trans. Commun.}, vol.~69,
  no.~8, pp. 5573--5592, 2021.

\bibitem{Emilpower_2020}
E.~Björnson and L.~Sanguinetti, ``Power scaling laws and near-field behaviors
  of massive {MIMO} and intelligent reflecting surfaces,'' \emph{IEEE Open J.
  Commun. Soc}, vol.~1, pp. 1306--1324, 2020.

\bibitem{Najafi_2021}
M.~Najafi, V.~Jamali, R.~Schober, and H.~V. Poor, ``Physics-based modeling and
  scalable optimization of large intelligent reflecting surfaces,'' \emph{IEEE
  Trans. Commun.}, vol.~69, no.~4, pp. 2673--2691, 2021.

\bibitem{Gradoni_2021}
G.~Gradoni and M.~Di~Renzo, ``End-to-end mutual coupling aware communication
  model for reconfigurable intelligent surfaces: An electromagnetic-compliant
  approach based on mutual impedances,'' \emph{IEEE Wireless Commun. Lett.},
  vol.~10, no.~5, pp. 938--942, 2021.

\bibitem{Qian_2021}
X.~Qian and M.~D. Renzo, ``Mutual coupling and unit cell aware optimization for
  reconfigurable intelligent surfaces,'' \emph{IEEE Wireless Commun. Lett.},
  vol.~10, no.~6, pp. 1183--1187, 2021.

\bibitem{Abrardo_2021}
\BIBentryALTinterwordspacing
A.~Abrardo, D.~Dardari, M.~D. Renzo, and X.~Qian, ``{MIMO} interference
  channels assisted by reconfigurable intelligent surfaces: {M}utual coupling
  aware sum-rate optimization based on a mutual impedance channel model,'' Feb.
  2021. [Online]. Available: \url{https://arxiv.org/abs/2102.07155}
\BIBentrySTDinterwordspacing

\bibitem{Sun1_2021}
Y.~Sun, C.-X. Wang, J.~Huang, and J.~Wang, ``A 3{D} non-stationary channel
  model for 6{G} wireless systems employing intelligent reflecting surface,''
  in \emph{Int. Conf. Wireless Commun. Signal Process. (WCSP)}, Nanjing, China,
  Oct. 2020, pp. 19--25.

\bibitem{Sun2_2021}
------, ``A 3{D} non-stationary channel model for 6{G} wireless systems
  employing intelligent reflecting surfaces with practical phase shifts,''
  \emph{IEEE Trans. Cogn. Commun.}, vol.~7, no.~2, pp. 496--510, 2021.

\bibitem{GSun_2021}
G.~{Sun \it{et al.}}, ``A 3{D} geometry-based non-stationary {MIMO} channel
  model for {RIS}-assisted communications,'' in \emph{IEEE Veh. Technol. Conf
  (VTC2021-Fall)}, Norman, OK, USA, 2021, pp. 1--5.

\bibitem{Jiang_2021}
H.~{Jiang \it{et al.}}, ``A general wideband non-stationary stochastic channel
  model for intelligent reflecting surface-assisted {MIMO} communications,''
  \emph{IEEE Trans. Wireless Commun.}, vol.~20, no.~8, pp. 5314--5328, 2021.

\bibitem{Xiong_2021}
B.~{Xiong \it{et al}}, ``A statistical mimo channel model for reconfigurable
  intelligent surface assisted wireless communications,'' \emph{IEEE Trans.
  Commun. (Earl Access)}, pp. 1--1, 2021.

\bibitem{SimRIS_Latincom}
E.~{Basar} and I.~{Yildirim}, ``{SimRIS} channel simulator for reconfigurable
  intelligent surface-empowered {mmWave} communication systems,'' in
  \emph{Proc. IEEE Latin-American Conf. Commun. (LATINCOM 2020)}, Nov. 2020,
  pp. 1--6.

\bibitem{SimRIS_TCOM}
E.~Basar, I.~Yildirim, and F.~Kilinc, ``Indoor and outdoor physical channel
  modeling and efficient positioning for reconfigurable intelligent surfaces in
  {mmWave} bands,'' \emph{IEEE Trans. Commun.}, vol.~69, no.~12, pp.
  8600--8611, 2021.

\bibitem{Kilinc_2021}
\BIBentryALTinterwordspacing
F.~Kilinc, I.~Yildirim, and E.~Basar, ``Physical channel modeling for
  {RIS}-empowered wireless networks in sub-6 {GHz} bands,'' Nov. 2021.
  [Online]. Available: \url{https://arxiv.org/abs/2111.01537}
\BIBentrySTDinterwordspacing

\bibitem{3GPP_5G}
``3{GPP} {TR} 38.901 {V}16.1.0 - {S}tudy on channel model for frequencies from
  0.5 to 100 {GH}z,'' Dec. 2019.

\bibitem{Nayeri}
P.~Nayeri, F.~Yang, and A.~Z. Elsherbeni, \emph{Reflectarray Antennas: Theory,
  Designs, and Applications}.\hskip 1em plus 0.5em minus 0.4em\relax USA:
  Wiley, 2018.

\bibitem{5G_Channel}
N.~Docomo, ``5g channel model for bands up to100 ghz,'' Tech. Report, Oct,
  Tech. Rep., 2016.

\bibitem{3GPP_Sub6G}
``3{GPP} {TR} 36.873 {V}12.7.0 - {S}tudy on {3D} channel model for {LTE},''
  Dec. 2017.

\bibitem{article}
P.~{Kyösti \it{et al.}}, ``{WINNER II} {c}hannel {m}odels,''
  \emph{IST-4-027756 WINNER II D1.1.2 V1.2}, Feb. 2008.

\bibitem{Pinv_2019}
\BIBentryALTinterwordspacing
T.~{Hou \it{et al.}}, ``{MIMO} assisted networks relying on large intelligent
  surfaces: {A} stochastic geometry model,'' Oct. 2019. [Online]. Available:
  \url{https://arxiv.org/abs/1910.00959}
\BIBentrySTDinterwordspacing

\end{thebibliography}

\end{document}